\documentclass[aps,twocolumn,showpacs]{revtex4}
\usepackage{graphics}
\usepackage{graphicx}

\def\ss{\scriptscriptstyle }
\begin{document}
\title{Voltage and temperature dependencies of conductivity in gated graphene}
\author{F.T. Vasko$^{1,2}$}
\email{ftvasko@yahoo.com}
\author{V. Ryzhii$^{2,3}$}
\affiliation{$^{1}$~Institute of Semiconductor Physics, NAS of Ukraine,
Pr. Nauki 41, Kiev, 03028, Ukraine \\
$^{2}$~University of Aizu, Ikki-machi, Aizu-Wakamatsu 965-8580, Japan \\
$^{3}$~Japan Science and Technology Agency, CREST, Tokyo 107-0075, Japan}
\date{\today}

\begin{abstract}
The resistivity of gated graphene is studied taking into account electron and
hole scattering by short- and long-range structural imperfections (the characteristics 
of disorder were taken from the scanning tunneling microscopy data) and by acoustic 
phonons. The calculations are based on the quasiclassical kinetic equation with the 
normalization condition fixed by surface charge. The gate-voltage and temperature 
effects on the resistance peak, which is centered at the point of intrinsic 
conductivity, are found to be in agreement with the transport measurements.
\end{abstract}


\maketitle
Numerous features of electronic properties of graphene (see discussion and Refs. 
in 1 and 2) are caused by both a neutrino-like dynamics of carriers \cite{2a}, which is 
described by the Weyl-Wallace model \cite{3}, and a substantial modification of scattering 
processes. For instance, the gapless energy spectrum and specific character of scattering 
are responsible for the peak of resistance of the gated graphene sheet (Fig. 1$a$). Such 
a peak was observed in Refs. 1 and 5-9 and one appears due to the transformation between 
$n$- and $p$-types of conductivity through the intrinsic region with the chemical potential 
in the vicinity of the band cross-point. Moreover anomalous temperature dependence was 
found: the peak resistance decreases with temperature while the tail resistance increases. 
To the best of our knowledge, a qualitative description of the gate-voltage and temperature 
effects on conductivity is not performed yet. 
Much attention was attached to the minimal conductivity phenomena, see last papers, 
\cite{7} and the metallic regime of conductivity was also examined.  
\cite{8} Recently, the near-maximum shape of the graphene resistance was considered 
assuming the scattering by a remote charge impurities layer is a dominant scattering 
mechanism. \cite{6} Since the high-density layer is not likely to be present, an additional 
scattering due to disorder was suggested in Ref. 12, but the only metallic regime of 
conductivity was discussed.

In this Brief Report, we calculate the gate-voltage and temperature dependencies of 
the resistivity peak taking into account structural inhomogeneities of graphehe [both 
long-range inhomogeneities and point defects, Figs. 1$b$ and 1$c$, respectively] and the 
acoustic phonon
scattering. Our model is justified by the recent scanning tunneling microscopy (STM) 
measurements, \cite{10} where both the imperfections with the lateral scale of 5$-$9 nm 
and point defects were reported. The long-range disorder ($DL$) is described by the
potential $U_{\bf x}$ with the Gaussian correlation function $\langle U_{\bf x}U_{\bf x'} 
\rangle \equiv\overline{U_l}^2\exp\{-[({\bf x}-{\bf x'})/l_c]^2\}$ where $\overline{U_l}$
is the averaged energy and $l_c$ is the correlation length. The short-range defects ($DS$) 
of sheet density $n_d$ are approximated by the potential $U_d\Delta ({\bf x})$ where 
$\Delta ({\bf x})$ is a function localized over the scale $l_d^2$. The main contribution 
to the acoustic phonon scattering appears due to the deformation interaction with 
longitudinal vibrations, \cite{11,11a} $D\nabla\cdot{\bf u}_{\bf x}$, where $D$ is the 
deformation potential and ${\bf u}_{\bf x}$ is the displacement vector of LA-mode.
\begin{figure}[ht]
\begin{center}
\includegraphics{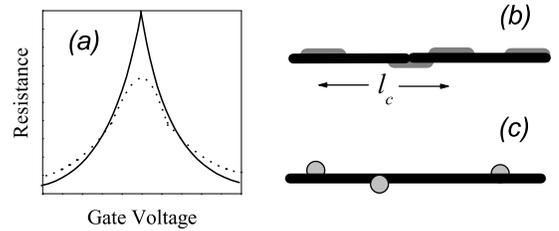}
\end{center}\addvspace{-1 cm}
\caption{($a$) Schematic plot of the resistance at zero (solid) and finite (dotted) 
temperatures versus gate voltage. Nonideal sheet of graphene ($b$) with a long-range 
inhomogeneities (gray) and ($c$) with point defects marked by balls.}
\end{figure}

The peak of resistance is originated from the distinctions of scattering
caused by $DS$ (or $LA$) and $DL$ mechanisms. Since a short-range relaxation rate (due
to $DS$ contribution or $LA$ phonon scattering at high temperatures) is proportional to 
the density of states \cite{8}, the resistance $R$ does not change with concentration
or gate voltage $V_g$. Besides, the $LA$ contribution is proportional to temperature
of phonons $T$, and $R$ increases with $T$. In contrast, the $DL$ scattering becomes
suppressed if an electron (hole) momentum exceeds $\hbar /l_c$. Due to this, $R$
decreases when $V_g$ or $T$ (i.e. energy of carriers) increases. Thus, a combination of 
$DL$- and $DS$- ($LA$-) mechanisms allows an explanation of an experimental behaviour 
shown in Fig.1$a$.

Scattering between electron or hole states \cite{12} with two-dimensional momenta 
$\bf p$ and $\bf p'$ is described by the transition probabilities $W_{{\bf p},
{\bf p}'}^{\ss (j)}$ where $j=DL$, $DS$, and $LA$. Within the Born approximation, 
$W_{{\bf p},{\bf p}'}^{\ss (j)}$ is written in the golden rule form,
%1
\begin{equation}
W_{{\bf p},{\bf p}'}^{\ss (j)}=\frac{{2\pi }}{\hbar }W_q^{\ss (j)}\Psi_{\theta }
\delta [v_{\ss W} (p - p')] ,
\end{equation}
where the overlap factor $\Psi_{\theta }=(1+\cos\theta )/2$ with $\theta =
\widehat{\mathbf{p'},\mathbf{p}}$ describes the suppression of backscattering 
processes, \cite{13} $v_{\ss W}\simeq 10^8$ cm/s is the characteristic velocity 
of the linear dispersion laws, $\pm v_{\ss W}p$, with the upper and lower signs 
corresponding to electrons and holes, respectively. The matrix element $W_q^{\ss (j)}$, 
which depends on the momentum transfer $\hbar{\bf q}=\mathbf{p}-\mathbf{p'}$, is given by
%2
\begin{equation} 
W_q^{\ss (j)}= \left\{ {\begin{array}{ll} 
\overline{U}_l^2\pi l_c^2e^{-(ql_c/2)^2} & (DL) \\ ~~~ n_dU_d^2\pi^2l_d^4 & (DS) \\ 
~~ {|C_q |^2 }2T/\hbar\omega_q & (LA).  \end{array}} \right. 
\end{equation}
Here the $DL$ and $DS$ mechanisms of scattering by static disorder are described by 
the Fourier transformations of $\langle U_{\bf x}U_{\bf x'}\rangle$ and $U_d\Delta (
{\bf x})$. The $LA$ probability is expressed via the matrix element $|C_q|^2=D^2\hbar 
\omega_q/(2\rho_{s}s^2L^2)$, where $s$ is the sound velocity, $\omega_q=sq$ is 
the phonon frequency, $\rho_{s}$ is the sheet density of graphene, and $L^2$ is the 
normalization area. Within the equipartition condition, at $\hbar\omega_q\ll T$, we 
replace the Planck distribution of phonons by $T/\hbar\omega_q$.
 
Under the steady-state electric field, $\bf E$, the electron ($e$) and hole ($h$) 
distribution functions, $f_{e\mathbf{p}}$ and $f_{h\mathbf{p}}$, are governed by the 
quasiclassical kinetic equations (here and below $k=e,h$),
%3
\begin{equation}
e\mathbf{E}\cdot \frac{\partial f_{k\mathbf{p}}}{\partial \mathbf{p}}
=\sum_{j\mathbf{p'}}W_{{\bf p},{\bf p'}}^{\ss (j)}(f_{k\mathbf{p'}}-
f_{k\mathbf{p}}) ,
\end{equation}
with the standard collision integral for the elastic scattering. \cite{14}
Using the electron and hole velocities $\pm v_{\ss W}{\bf p}/p$, one obtains the 
following expression for the current density:
%4
\begin{equation}
\mathbf{I}=\frac{4ev_{\ss W}}{L^2}\sum_{\mathbf{p}}\frac{\mathbf{p}}p\left(
f_{e\mathbf{p}}+f_{h\mathbf{p}}\right) ,
\end{equation}
where factor 4 appears due to the spin and valley degeneracy. The distribution
functions $f_{e\mathbf{p}}$ and $f_{h\mathbf{p}}$ are also related to each other 
and to the sheet charge density, $Q_s$,
%5
\begin{equation}
Q_s=\frac{4e}{L^2}\sum_{\mathbf{p}}\left( f_{e\mathbf{p}}-f_{h\mathbf{p}%
}\right) .
\end{equation}
Note, that $Q_s$ can be expressed via the gate voltage as $V_g=\epsilon Q_s/4\pi d$.
Below we assume the SiO$_2$ substrate of the width $d=$3 nm with the dielectric permittivity
$\epsilon\simeq$3.

Within the linear approximation, the solution of Eq. (3) can be seached in the form 
$f_{k\mathbf{p}}=F_{kp}+\Delta f_{k\mathbf{p}}$, where $F_{kp}=\{\exp [(v_{\ss W} p\mp\mu)/T]
+1\}^{-1}$ are the equilibrium Fermi distributions with the chemical potential $\mu$. 
The asymmetric parts of distribution functions, $\Delta f_{k\mathbf{p}}$, can be 
obtained from Eq. (3) in the form
%6
\begin{equation}
\Delta f_{k\mathbf{p}}=-\frac{(e\mathbf{E}\cdot\mathbf{p})}{p}\tau_p^{\ss (m)}
\left( -\frac{dF_{kp}}{dp}\right) , 
\end{equation}
with the momentum relaxation time, $\tau_p^{\ss (m)}$, is given by the sum of relaxation
rates $\tau_p^{\ss (m)}=\left[\sum_j(1/\tau_p^{\ss (j)})\right]^{-1}$ where $1/\tau_p^{\ss (j)}
=\sum_{\mathbf{p'}}W_{{\bf p},{\bf p}'}^{\ss (j)}(1-\cos\theta )$. Substituting Eq. (6) 
into Eq. (4) one obtains the resistivity, which is introduced by ${\bf I}={\bf E}/R$, 
as follows:
%7
\begin{equation}
\frac{1}{R} =\frac{e^2v_{\ss W}}{\pi\hbar^2}\int_0^{\infty}dp~p\tau_p^{\ss (m)}
\sum_{k}\left( -\frac{dF_{kp}}{dp}\right)  
\end{equation}
and the momentum relaxation time here do not depend on $k$. It is convenient to present
the relaxation rates as
%8
\begin{equation}
\frac{1}{\tau_p^{\ss (j)}}=\frac{v_{sc}^{\ss (j)}p}{\hbar}\left\{\begin{array}{ll}
\Psi\left(pl_c/\hbar\right)  & ~~~(DL) \\ ~~~~1  & (DS, ~LA) ,  \end{array} \right. 
\end{equation}
where the averaging over angle gives the function 
%9
\begin{equation}
\Psi(z)=\frac{e^{-z^2/2}}{z^2}I_1\left(\frac{z^2}{2}\right) ,
\end{equation}
with the first-order Bessel function of an imaginary argument, $I_1(x)$. The characteristic 
velocities, $v_{sc}^{\ss (j)}$, in Eq. (8) are introduced as follows:
%10
\begin{equation}
v_{sc}^{\ss (j)}=\frac{1}{4\hbar^2v_{\ss W}}\left\{\begin{array}{ll} \pi\overline{U}_l^2
l_c^2  & (DL) \\ \pi^2U_d^2l_d^3n_d  & (DS) \\ D^2T /(\rho_ss^2)  & (LA) ,
\end{array} \right.  
\end{equation}
so that $v_{sc}^{\ss (j)}\ll v_{\ss W}$ for the parameters used below.

Next, we simplify the expression (7) using the dimensionless resistivity, 
$Re^2/\pi\hbar$. We consider the case of degenerate carriers, $|\mu |\gg T$, 
when Eq. (5) gives the square-root dependence 
of $\mu$ on the gate-voltage: $\mu =\hbar v_{\ss W}\sqrt{\epsilon V_g/4|e|d}$. 
Since $\sum_{k}(-dF_{kp}/dp)$ can be replaced by $\delta$-function, Eqs. (7)-(9) 
yield the dimensionless resistivity in the following form:
%11
\begin{equation}
R \frac{e^2}{\pi\hbar}=\frac{v_{sc}^{\ss (DL)}}{v_{\ss W}}\Psi\left(\frac{\mu l_c}
{v_{\ss W}\hbar}\right)+\frac{v_{sc}^{\ss (DS)}}{v_{\ss W}}+\frac{v_{sc}^{\ss (LA)}}
{v_{\ss W}} ,
\end{equation}
with the temperature dependent $LA$ contribution $v_{sc}^{\ss (LA)}/v_{\ss W}$. Another
simple expression can be obtained for the case of  the intrinsic regime of transport, 
$Q_s=0$, when $\mu$=0 and $F_{kp}\rightarrow [\exp (v_{\ss W} p/T)+1]^{-1}$ in Eq.(7).
As a result, the inverse sheet resistance is expressed through the dimensionless integral,
%12
\begin{eqnarray}
\frac{\pi\hbar}{Re^2}=\frac{v_{sc}^{\ss (DL)}}{v_{\ss W}}\int_0^{\infty}
\frac{dx}{1+\cosh x}  ~~~\\
\times \left[\Psi\left( x\frac{Tl_c}{v_{\ss W}\hbar}\right)+
\frac{v_{sc}^{\ss (DS)}}{v_{\ss W}}+\frac{v_{sc}^{\ss (LA)}}{v_{\ss W}}
\right]^{-1} . \nonumber
\end{eqnarray} 
Here, not only the ratio $v_{sc}^{\ss (LA)}/v_{\ss W}$ increases with the temperature but 
also the $DL$ contribution becomes suppressed, if $Tl_c /v_{\ss W}\hbar >1$.
\begin{figure}[ht]
\begin{center}
\includegraphics{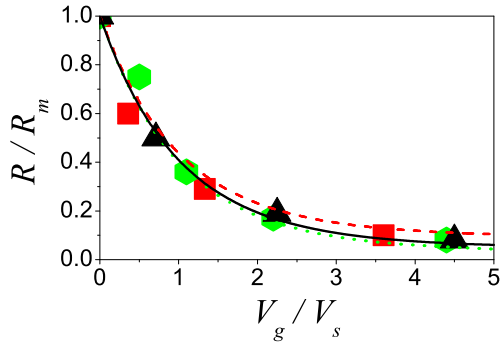}
\end{center}
\addvspace{-1 cm}
\caption{(Color online) Normalized resistance $R/R_m$ versus dimensionless gate voltage 
$V_g/V_s$ plotted for parameters of ref. 1 (dotted curve, hexagons), Ref. 7 (solid curve, 
triangles) and Ref. 8 (dashed curve, squares).}
\end{figure}

We turn now to discussion of the experimental data using the parameters of Eq. (2) in
order to fit the gate-voltage and temperature dependencies of resistance obtained in
Refs. 1, 7, and 8 (similar low-temperature results can be obtained from Refs. 6 and 9). 
Under the description of the $LA$ scattering we use the known parameters $s\simeq 7.3\cdot 
10^5$ cm/s and $\rho_s\simeq 7\cdot 10^{-8}$ g/cm$^2$. In order to 
obtain the temperature-dependent contribution to the tails of peak [about 100 $\Omega$ 
in Fig.2$a$ of Ref. 7 and in Fig.1 of Ref.8] one needs the deformation potential 
$D\simeq$12 eV which is about 75$\%$ of the graphite value. \cite{11,13} For such a set 
of parameters, one obtains $v_{sc}^{\ss (LA)}\simeq 8\cdot 10^5$ cm/s at $T$=300 K from 
Eq. (11). Since $\Psi (z\rightarrow\infty)=0$, the temperature independent part of the 
tails, which is about 400 $\Omega$ in Fig. 1 of Refs. 1 and 8 and about 200 $\Omega$ in 
Fig. 2$a$ of Ref. 7, appears due to the $DS$ contribution with the characteristic 
velocities $v_{sc}^{\ss (DS)}\simeq 3.2\cdot 10^6$cm/s and $1.6\cdot 10^6$cm/s for 
Refs. 1 and 8 and Ref. 7, respectively.
\begin{figure}[ht]
\begin{center}
\includegraphics{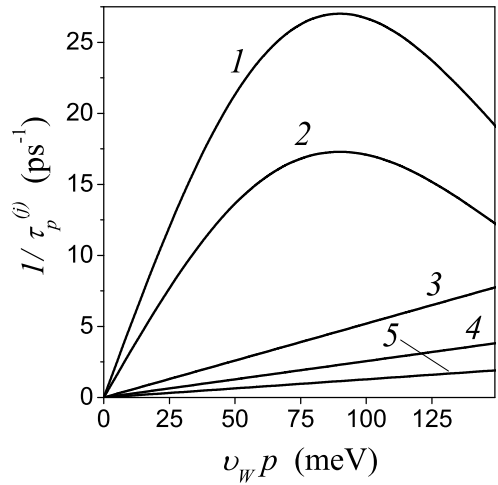}
\end{center}
\addvspace{-1 cm}
\caption{Relaxation rates $1/\tau_p^{\ss (j)}$ versus energy, $v_{\ss W}p$: ($1$) $j=DL$ 
for parameters of Refs. 7 and 8, ($2$) $j=DL$ for Ref.1, ($3$) $j=DS$ for Refs. 1 and 7, 
($4$) $j=DS$ for Ref. 8, and ($5$) $j=LA$ for Refs. 1, 7, and 8.}
\end{figure}

In order to fit the gate-bias dependencies, in Fig. 2 we plot the normalized resistance
$R/R_m$ versus the dimensionless gate-bias, $V_g/V_s$, where $R_m=\pi\hbar (v_{sc}^{\ss (DL)}
+v_{sc}^{\ss (DS)})/v_{\ss W}e^2$ and $V_s=4|e|d/\epsilon l_c^2$. Here we used the  
values of $v_{sc}^{\ss (DS)}$ obtained above and the 5 nm correlation length, which is taken 
from the STM measurements \cite{10}, for all the cases. \cite{1,4,5} A good agreement is 
obtained for $v_{sc}^{\ss (DL)}\simeq 3.2\cdot 10^7$cm/s (Refs. 7 and 8) and $5\cdot 10^7$cm/s. 
\cite{1} Thus, we have found the scattering parameters of the relaxation rates given by Eqs. 
(8) and (9) for three different samples and one can plot $\nu_p^{\ss (j)}$ versus energy 
$v_{\ss W}p$ (see Fig. 3). The velocities $\nu_p^{\ss (DL,DS)}$ are connected to the 
characteristic potentials $\overline{U}_l$ and $U_d$ as follows. For the long-range disorder, 
we obtain $\overline{U}_l \sim$100 meV or 80 meV for Ref. 1 or Refs. 7 and 8, respectively, 
i.e. $\overline{U}_l$ is less than the interlayer coupling energy $\sim$350 meV. The energy 
$U_d$ can be estimated as $\sim$3 eV (of the order of the intralayer coupling energy), if we 
use $l_d\simeq$2 nm and $n_dl_d^2\sim 1\%$. The suppression of resistance with $V_g$ takes 
place if the chemical potential exceeds the 100 meV threshold, see Fig. 3 ($\mu\sim$50 meV at 
$V_g$=10 V for the structures under consideration). Thus, both voltage-induced and  
temperature-induced (below 700$-$1000 K where the optical phonon contribution becomes 
essential) suppression of resistance takes place.
\begin{figure}[ht]
\begin{center}
\includegraphics{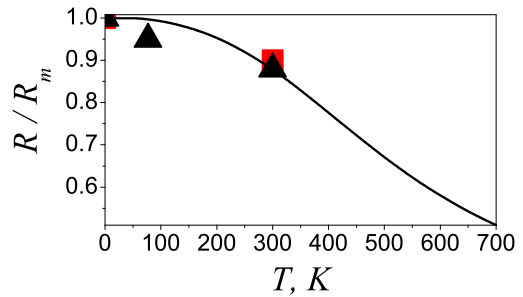}
\end{center}\addvspace{-1 cm}
\caption{(Color online) Temperature dependence of resistance $R/R_m$ for the intrinsic 
regime of conductivity, $\mu =0$. The experimental data taken from Refs. 7 and 8 are 
marked as in Fig. 2.}
\end{figure}

Due to lack of data on the temperature dependency of resistance, we only discuss here the
suppression of the resistance peak value when temperature increases. We use Eq. (12) 
which governs the intrinsic regime of conductivity and the dependency $R(T)$ is
in agreement with the room temperature data of Refs. 7 and 8 without any additional parameters. 
Figure 4 shows an essential suppression of $R$ when the ratio $Tl_c/v_{\ss W}\hbar$ exceeds 
unit (at $T>$300 K for the parameters used). An extrapolation for the high-temperature region 
shows an essential suppression of the $DL$ scattering at $T\geq$1000 K. Thus, a minimum of 
resistance, which is caused by such a suppression and an optical phonon contribution, should 
take place for the temperature region above $\sim$ 500 K.

Let us discuss the approximations used in our calculations. The main assumption of 
our model is the phenomenological description of the elastic scattering by long- and
short-range structural inhomogeneities. While the in-plane scales of disorder, $l_c$ and 
$l_d$, are clear and justified by Ref. 13, a nature of inhomogeneities and a value of the
 phenomenological energies [$\overline{U}_l$ and $U_d$ in Eq.(2)] require an additional
analysis. Other approximations made are rather standard and generally accepted.  
We have neglected the screening effects because it is not a leading effect for 
a spatial ranges up to 5 nm. The only scattering by $LA$ mode is taken into account because 
another contributions are weaker (see Refs. 14 and 17 and references therein). 
Since $v_{sc}^{\ss (j)}\ll v_{\ss W}$, the (anti)localization phenomena \cite{16} can 
be neglected if $Re^2/\pi\hbar\ll 1$ (but it may be essential in Ref. 5 and in Fig. 3 of of 
Ref. 7, where $R_m\sim$ 10 K$\Omega$). In addition, the quantum dynamics restrictions, like a 
Zitterbewegung effect \cite{17} or the field-induced interband tunneling, are only essential 
in the vicinity of the cross-point. Since $R$ is weakly changed near this point, one can use 
the results obtained as phenomenological dependencies. However, a justification for the scales 
$\sim$1 V or tens of kelvins is beyond of the consideration performed.
 
In summary, we have found that the main factor, which forms the resistance peak centered 
at the intrinsic conductivity region, is a structure disorder (e.g., defect clusters, bilayer 
islands, or a substrate disorder \cite{18}) with the scale around 5 nm for the samples 
analyzed. Short-range defect
contribution and acoustic phonon scattering are only essential for the heavily doped 
case on the tails of peak. Both gate-voltage and temperature-induced quenchings of
the resistance peak are in agreement with the transport measurements \cite{1,4,5}
and with the STM mapping of nonideal graphene sheet. \cite{10} A complete verification of
the mechanism suggested requires both microscopic calculations (in order to estimate 
the fitting parameters $\overline{U_l}$ and $U_d$) and a further investigation of the 
transport phenomena. Particularly, high-temperature conductivity measurements, which can
confirm a suppression of resistance peak, as well as treatment of magnetotransport 
and high-frequency responses are necessary.  

To conclude, the results presented permit to analyze linear characteristics of
the graphene-based field-effect transistor. Recent studies of such a device, see 
Ref. 22 and references therein, demonstrated an essential gate modulation. 
However, the data available are not sufficient to determine the scattering mechanisms. 
We believe that t result obtained will stimulate further investigations in order to 
understand transport phenomena and to improve device characteristics. 

The work at the University of Aizu was partially supported by the Grant-in-Aid for 
Scientific Research (S) from the Japan Society for Promotion of Science and by the
Japan Science and Technology Agency, CREST.


\begin{thebibliography}{12}
\bibitem{12} The interband transitions are forbidden both for static disorder
(due to energy conservation) and for LA-mode (due to the condition $s<v_{\ss W}$).
\end{thebibliography}
\end{document}